\documentclass{PoS}

\title{Non-perturbative calculation of $Z_m$  using \\Asqtad fermions}

\ShortTitle{Non-perturbative calculation of $Z_m$ using Asqtad fermions}

\author{\speaker{Andrew T. Lytle}\\
       Physics Department, University of Washington, Seattle, WA 98195-1560 \\
       E-mail: \email{atlytle@gmail.com}}

\abstract{We report progress on a non-perturbative
calculation of the light quark mass renormalization factor $Z_m^
{\overline{MS},LAT}(\mu,a)$, using dynamical Asqtad fermions.  This
quantity is used to determine the light quark masses in the
conventional $\overline{MS}$ scheme.   Such a non-perturbative
determination of $Z_m$ removes uncertainties due to truncation of its
perturbative expansion, currently known to two loops.  These
calculations have been carried out  using publicly available MILC
lattices with \break spacings of $\approx 0.125$ and $0.09$ fm.}

\FullConference{The XXVII International Symposium on Lattice Field Theory\\
                July 26-31, 2009\\
                Peking University, Beijing, China}

\begin{document}

\section{Introduction}
Quark masses are fundamental parameters of the standard model, and as such are inputs whose determination must ultimately come from experimental measurement.  Lattice simulations of QCD provide a means by which these parameters may be precisely determined and their values systematically improved upon.  Because quark masses are scheme and scale dependent, one must determine ``$Z$-factors'' to convert the bare lattice masses determined from simulations to a continuum scheme.

$Z$-factors may be determined using lattice perturbation theory, but these determinations necessarily suffer from uncertainties due to truncation of the perturbative expansion.  Extending these calculations to higher order is difficult, particularly when using improved actions as is done in most current simulations.  Another strategy, more amenable to systematic improvement, is to compute $Z$-factors non-perturbatively using the lattice simulation itself.

In this article we report on progress in this area using staggered quarks and the Asqtad~\cite{Orginos99}~\cite{Lepage98}~\cite{Lepage92}  $O(a^4, a^2 \alpha_s)$-improved action.  This is the first such unquenched study using staggered quarks.  First we describe the methodology of non-perturbative renormalization (NPR)~\cite{Martinelli94} used in the study, highlighting details specific to the use of staggered fermions.  Then we present our results, and conclude with a discussion of further work to be done.
\section{Methodology}
\subsection{Basic principles}
Quark masses are scheme and scale dependent quantities.  For a given lattice regularization, one tunes the bare input masses to the physical point of QCD.  To compare amongst different determinations, these quantities must be converted to a common scale and scheme.  This is conventionally taken to be the $\overline{MS}$ scheme at a scale of $\mu = 2 GeV$:
\begin{equation}
m^{\overline{MS}}(\mu) = Z_m^{\overline{MS},LAT}(\mu,a) \,\, \frac{am_0}{a}.
\end{equation}
The factor $Z_m^{\overline{MS},LAT}(\mu,a)$ has been calculated for the Asqtad action to two loops using lattice perturbation theory~\cite{Mason05}.

Because the $\overline{MS}$ scheme is a continuum scheme, one must first convert to a scheme that is sensible in both lattice and continuum theories, and then use an additional matching factor to convert between schemes.  We take the intermediate scheme to be the $RI/MOM$ 
scheme~\footnote{In practice we use the $RI'$ scheme.  This differs from the $RI$ scheme in the definition of the wavefunction renormalization factor $Z_q$~(\ref{eq: Zq}) and results in a difference of $O(1\%)$~\cite{Franco98}.}, to be explained momentarily:
\begin{equation}
Z_m^{\overline{MS},LAT}(\mu,a) = 
Z_m^{\overline{MS},RI}(\mu) \,\, U^{RI}(\mu,p) \,\, Z_m^{RI,LAT}(p,a).
\end{equation}
Here $U^{RI}(\mu,p)$ is a perturbative running factor computed in the continuum, known to 4 loops, and $Z_m^{\overline{MS},RI}(\mu)$ is the $\overline{MS}$ to $RI$ conversion factor,  calculated to 3 loops in the continuum~\cite{Chetyrkin99}.  If one was content to work in the $RI$ scheme, this latter factor could be avoided entirely.

In the $RI/MOM$ scheme, $Z$-factors are defined such that Green's functions in a fixed (Landau) gauge, suitably projected, take their tree-level values.  Such $Z$-factors may be computed directly from lattice simulations \cite{Martinelli94}. One only requires that the scale of external momenta $p$ is chosen such that $ \Lambda_{QCD} \ll |p| \ll \frac{1}{a} $.  This ensures that non-perturbative contributions to $Z$-factors and lattice artifacts are simultaneously under control. Thus the NPR method trades the truncation uncertainty of a perturbative evaluation for uncertainties due to lattice artifacts, non-perturbative contributions, and statistics. These, however, are more amenable to systematic reduction.

In principle, one could extract $Z_m$ using only the following projections of the quark propagator:
\begin{equation} \label{eq: Zq}
Z_q(p) = -i \frac{1}{12 N_T} \sum_\mu
\frac{p_\mu}{p^2} 
\mbox{Tr}\left[\overline{\overline{(\gamma_\mu\otimes 1) }} S^{-1}[p]
\right],
\end{equation}
\begin{equation} \label{eq: ZqZm}
\frac{1}{12 N_T} \mbox{Tr}\left[\overline{\overline{(1\otimes 1)}} S^{-1}[p]\right]
= Z_q(p)
\left[ Z_m (p) m + C_1 \frac{\langle \bar q q\rangle}{p^2} + \dots\right].
\end{equation}
The propagator $S[p]$ is a 16-by-16 matrix in spin-taste space, as are the double-barred projection matrices. The latter are the staggered analogues of their continuum counterparts.
In practice, the projection (\ref{eq: ZqZm}) suffers severely from non-perturbative contamination due to the condensate term~\cite{Politzer76}.  One must make recourse to Ward identities to obviate this issue.

The lattice Ward identities of interest relate the scalar and pseudoscalar bilinear operators to the quark propagator:
\begin{equation} \label{eq: Zq/ZP}
\frac{1}{m}\mbox{Tr}\left[\overline{\overline{(1\otimes 1)}} S^{-1}[p]\right] = \mbox{Tr} 
\left[ 
\overline{\overline{(\gamma_5 \otimes \xi_5) }} \Lambda^{(\gamma_5 \otimes \xi_5)}[p]
\right] = \frac{Z_q[p]}{Z_P[p]},
\end{equation}
\begin{equation} \label{eq: Zq/ZS}
\frac{\partial}{\partial m}\mbox{Tr}\left[\overline{\overline{(1\otimes 1)}} S^{-1}[p]\right] = \mbox{Tr} 
\left[ 
\overline{\overline{(1 \otimes 1) }} \Lambda^{(1 \otimes 1)}[p]
\right] = \frac{Z_q[p]}{Z_S[p]}.
\end{equation}
Here the $\Lambda$ are amputated vertices of the bilinear operator with the indicated spin-taste structure.

In perturbation theory, one has the identity $Z_m = \frac{1}{Z_S} = \frac{1}{Z_P}$. On the lattice, the equality is spoiled by non-perturbative contributions, which can be seen from inserting 
(\ref{eq: ZqZm}) into (\ref{eq: Zq/ZP}) and (\ref{eq: Zq/ZS}).  One expects that the scalar bilinear in (\ref{eq: Zq/ZS}) gives a closer approximation to $Z_m$ because the derivative $\frac{\partial}{\partial m}$ will eliminate the $\frac{1}{p^2}$ condensate pole.

\subsection{Implementation}
Computations are being carried out using the Chroma \cite{Edwards03} software library for Lattice QCD.  We have added support for staggered momentum-source inversions and written routines to compute the bilinears of interest.  All configurations are first fixed to Landau gauge.
For a given momentum $p'$ in the reduced Brillouin zone $(0 \le p'_\mu < \frac{\pi}{a})$, 16 inversions are carried out on momentum sources to produce the correlators:
\begin{equation}
< \chi(x) \bar{\chi}(-p' + \pi B) > \, \equiv \, < \chi(x) \bar{\phi}_B(-p') >.
\end{equation}
Here $B$ is one of the sixteen binary vectors e.g. $(1,1,0,0)$.  Note that these momentum sources, being simple phase factors of the form $e^{ip \cdot x}$, differ from those used in the quenched study~\cite{Aoki99}, and result in a propagator that does not explicitly break taste symmetry.
From these correlators the ``polespace'' propagators are constructed by performing sixteen Fourier transforms on the free space index:
\begin{equation}
S_{AB}[p'] = \int d^4x \, e^{i(p'+\pi A)\cdot x} <\chi(x) \bar{\phi}_B(-p')>.
\end{equation}
Thus the propagators are 16-by-16 matrices in spin-taste space.

Inversions are carried out on publicly available MILC lattices \cite{Aubin04} with 2+1 flavors of Asqtad quarks, and are computed on Fermilab clusters using resources from a USQCD grant.  We use both fine ($a \approx 0.09$ fm) and coarse ($a \approx 0.125$ fm) lattices.  On the coarse lattices, we have varied both the sea and valence masses to produce unquenched data with masses $ am = .03,\,.02,\,.01$.
On the fine lattices we have studied masses $am_{val} = .018,\,.012,\,.006$ and 
$am_{sea} = .0062$, and intend to alter the sea quark masses to produce unquenched data.

We study the propagators and bilinears over a range of momenta to understand the non-perturbative and discretization effects.  The momenta are chosen using the naive "cylindrical cut" method \cite{deSoto07} to reduce discretization effects.  Data shown in this paper are all obtained using 8 lattices.  This results in relatively small statistical uncertainties because of our use of momentum sources.

\section{Results}
\begin{figure}[p]
\includegraphics[width=1.0\textwidth]{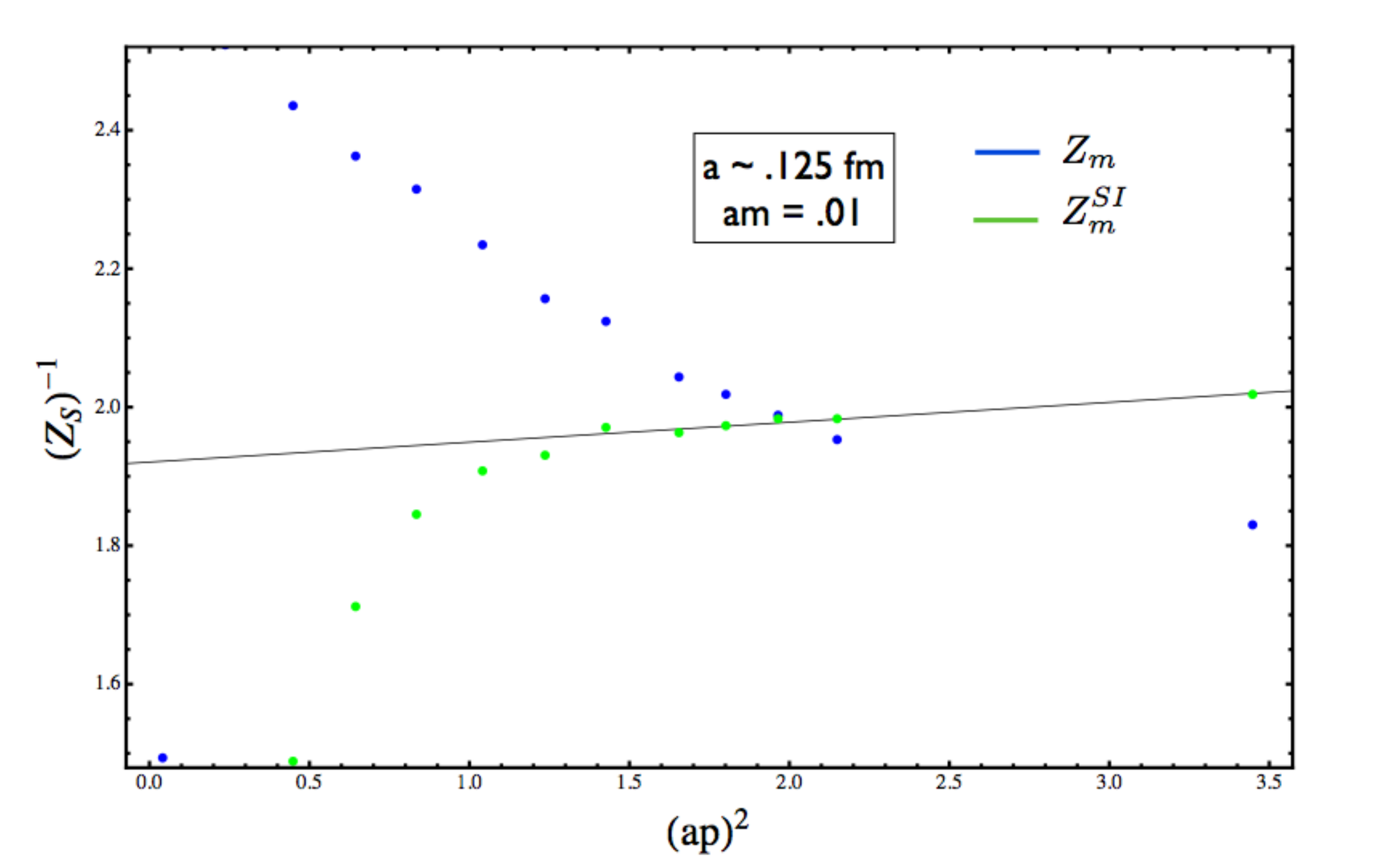}
\caption{Result for $Z_m = Z_S^{-1}$.  After applying the perturbative running factor to obtain $Z_m^{SI}$, the data noticeably flattens out and is consistent with a straight line at higher $(ap)^2$.}
 \label{fig: Zm} 
 \end{figure}

\begin{figure}[p]
\includegraphics[width=1.0\textwidth]{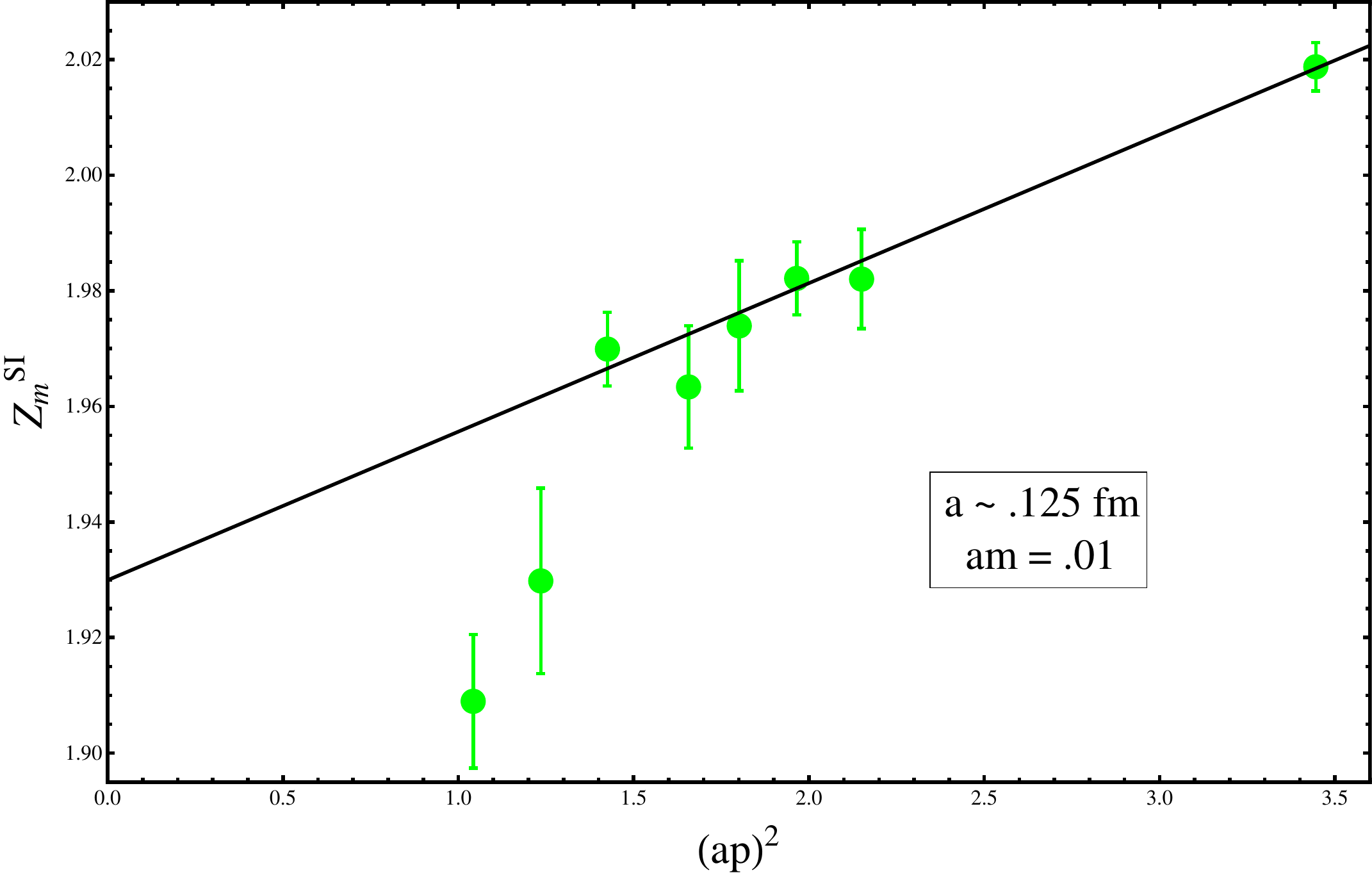}
\caption{Closeup of Figure 2 with error bars and $\chi^2$ fit to last 6 data pts.  Notice the small range on the y-axis.}
 \label{fig: Zm_close} 
 \end{figure}

\begin{figure}[p]
\includegraphics[width=1.0\textwidth]{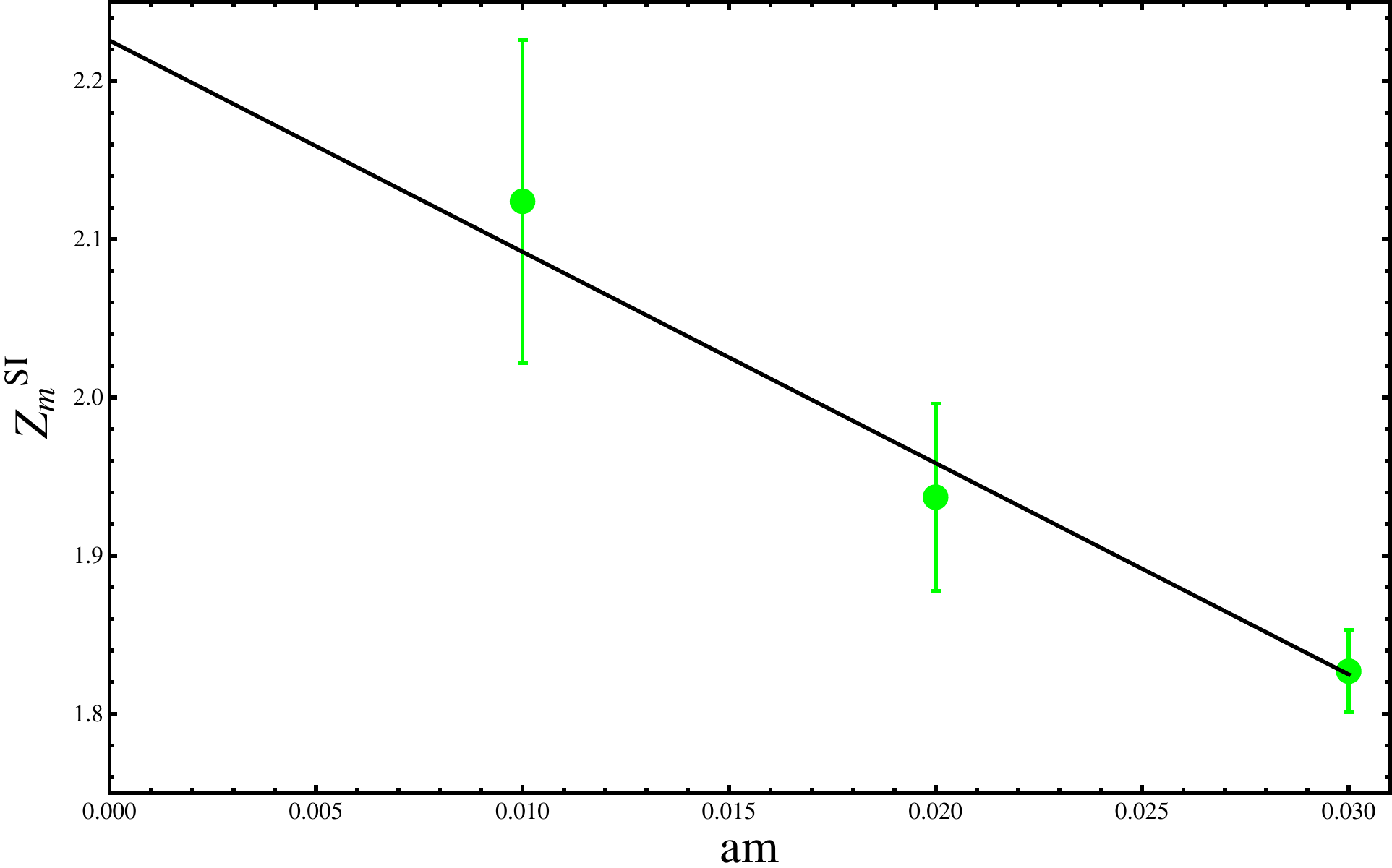}
\caption{Naive chiral extrapolation of $Z_m^{SI}$ on the coarse lattices (unquenched data).}
 \label{fig: chi_extrap} 
 \end{figure}

Figure~\ref{fig: Z_S,P} shows typical results for the scalar and pseudoscalar bilinears.  The $1/p^2$ non-perturbative contamination is clear in the graph of $Z_P^{-1}$, but is absent in $Z_S^{-1}$, consistent with the discussion in section 2.  
Figure~\ref{fig: Zm} shows a closeup of the same $Z_S^{-1}$ data, as well as the "scale-invariant" result, $Z_m^{SI}$, obtained using the perturbative running factor to run the data to a common scale, here $(ap)^2 = 2$.  If discretization effects were negligible, one would expect $Z_m^{SI}$ to be a constant, at least in the region of large $(ap)^2$ where the perturbative expansion is sensible and non-perturbative contributions are small.  Discretization effects proportional to $(ap)^2$ would result in the curve being a straight line with non-zero slope. The observed data is consistent with this.  Thus we remove artifacts by using the intercept of the interpolated line.  This is detailed in Figure~\ref{fig: Zm_close}.  The fitted points are chosen to minimize the $\chi^2$ per dof of a straight line fit.  
In Figure~\ref{fig: chi_extrap} we have plotted the result of such a procedure for the three masses on the coarse MILC lattices.  Also shown is a naive linear fit to the data, which, when combined with the bare strange mass and lattice spacing as determined by the MILC collaboration~\cite{Aubin04}, gives a final strange quark mass of $m_s^{\overline{MS}} = 106(6) MeV$ on the coarse lattices.  This is to be compared with the perturbative value of $m_s^{\overline{MS}} = 84(5) MeV$ obtained on the coarse lattices by the MILC collaboration~\cite{Mason05}.
 
 \begin{figure}[p]
\includegraphics[width=1.0\textwidth]{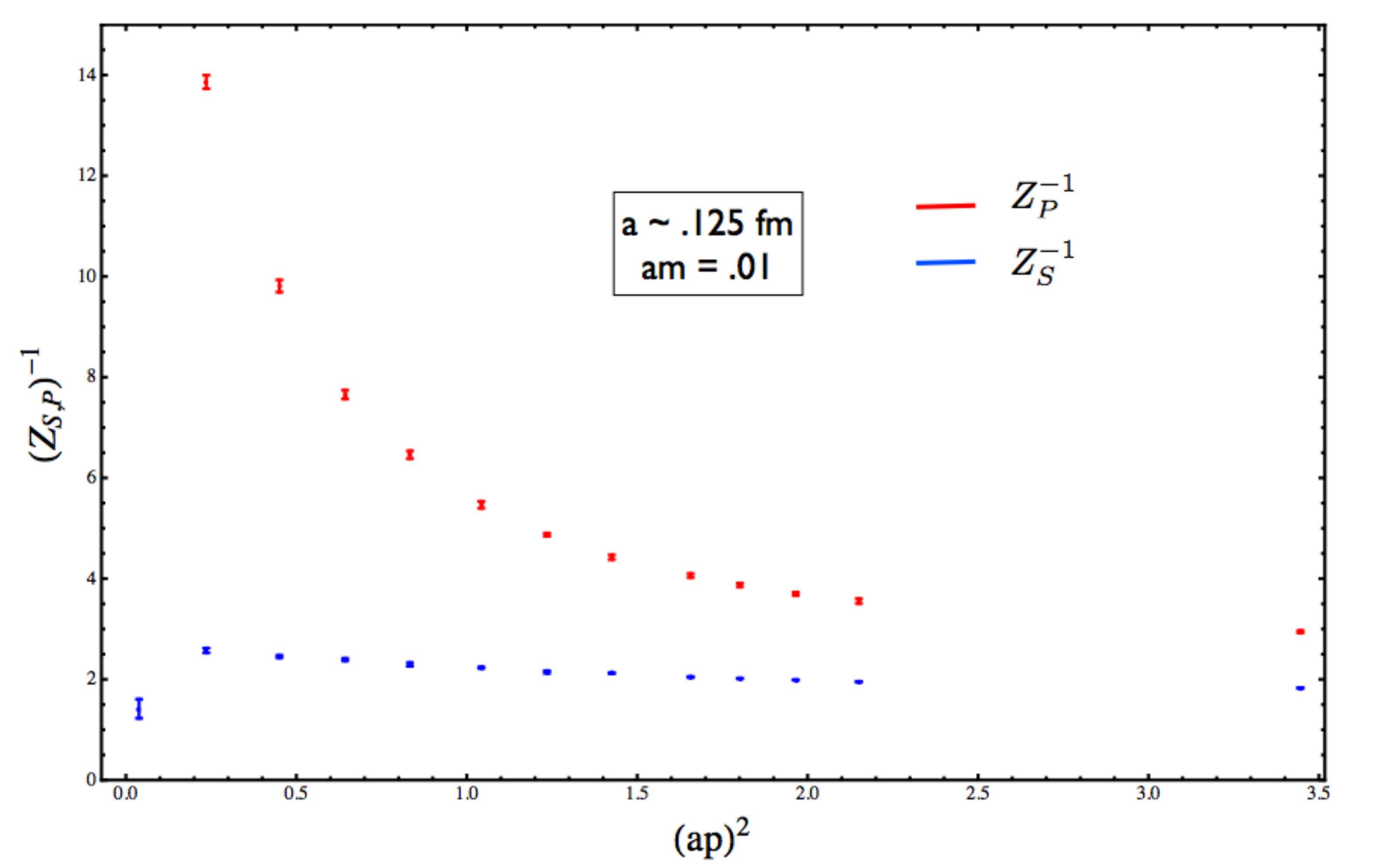}
\caption{$Z_P^{-1}$ and $Z_S^{-1}$ on the MILC coarse lattice ensemble (8 configs). The $1/p^2$ non-perturbative contamination in $Z_P^{-1}$ is clearly visible.}
 \label{fig: Z_S,P} 
 \end{figure}
 
\section{Conclusions, Outlook, Future Work}
It should be noted that the above analysis is preliminary.  A linear chiral extrapolation has been carried out, but a glance at Figure~\ref{fig: chi_extrap} shows that statistics must be improved to verify whether this is permissible.  In addition, analogous calculations must be performed on the fine lattices to permit a continuum extrapolation.  We also intend to calculate $Z_q$ using the vector and axial-vector bilinears, as this makes use of the $RI$ scheme directly and has no uncertainty associated with the choice of definition of $Z_q$~\cite{Blum01}.

Current data on the coarse lattices indicate a strange quark mass significantly larger than that obtained by perturbative analysis.  The fine lattice data also appear to support this result, though a fully unquenched analysis is pending.  This seems to be  consistent with a general trend that the non-perturbatively obtained masses are larger than those obtained perturbatively~\cite{Scholz09}~\cite{Aoki09}. This could be due to deficiencies of the perturbative expansions, or to systematic non-perturbative effects that have not been accounted for.  The use of non-exceptional momenta should help to distinguish these possibilities~\cite{Aoki09}.

\acknowledgments
I would like to thank S. Sharpe for useful conversations and support.  Computations for this work were carried out in part on facilities of the USQCD Collaboration, which are funded by the Office of Science of the U.S. Department of Energy.  This study was also supported in part by the U.S. Department of Energy under Grant No. DE-FG02-96ER40956.

\end{document}